\documentclass[]{spie}  

\usepackage{amsmath,amsfonts,amssymb}
\usepackage{graphicx}
\usepackage[colorlinks=true, allcolors=blue]{hyperref}
\usepackage{comment}
\usepackage{siunitx}

\title{Towards Autodifferentiable Point Spread Function Modelling for the VLT/ERIS Vortex Coronagraph}

\author[a]{P. Van de Walle Suárez}
\author[a]{G. Orban De Xivry}
\author[a]{O. Absil}
\author[b,c]{M. J. Bonse}
\author[d]{R. J. De Rosa}
\author[e,f,a]{V. Christiaens}

\affil[a]{STAR Institute, Universit\'e de  Li\`ege, all\'ee du Six Ao\^ut 19c, 4000 Li\`ege, Belgium}
\affil[b]{European Southern Observatory, Karl-Schwarzschild-Straße 2, 85748 Garching bei München, Germany}
\affil[c]{Max Planck Institute for Intelligent Systems, Max-Planck-Ring 4, 72076 Tübingen, Germany}
\affil[d]{European Southern Observatory, Alonso de Córdova 3107, Vitacura, Santiago, Chile}
\affil[e]{Université Paris-Saclay, Université Paris Cité, CEA, CNRS, AIM, F-91191 Gif-sur-Yvette, France}
\affil[f]{Université Paris-Saclay, CNRS, Institut d’Astrophysique Spatiale, 91405 Orsay, France}

\authorinfo{Further author information: (Send correspondence to P. Van de Walle Suárez)\\ E-mail: pvandewallesuarez@uliege.be}

\usepackage{aas_macros}

\begin{document} 
\maketitle

\begin{abstract}

High-contrast imaging of exoplanets is limited by the presence of quasi-static speckles and existing post-processing methods remain above the theoretical noise limit. We investigate a novel framework that incorporates adaptive optics (AO) wavefront sensor telemetry into post-processing using a differentiable optical model of the ERIS vortex coronagraph. As a first proof of concept, we inject AO telemetry into an ERIS forward model and assess its ability to reproduce the on-sky point spread function (PSF). We find that jitter discrepancies between the science camera and the AO telemetry hinder accurate PSF reconstruction. We proposed a plan to  characterise this jitter and find its mechanical origin. 

\end{abstract}


\section{INTRODUCTION}
\label{sec:intro}  

High-contrast imaging (HCI) of exoplanets requires image processing methods capable of 
obtaining direct images of planetary systems while overcoming the extreme contrast in brightness between the  faint planets and their parent star, as well as the residual starlight that remains in the images as quasi static artifacts known as speckles. Speckles are due to wavefront aberrations introduced in the science image by several factors such as, diffraction within the optical train, imperfect corrections from the adaptive optics (AO) system, and non-common path aberrations (NCPAs) between the wavefront sensor (WFS) and the science camera \cite{goebel2018measurements}. They evolve on a variety of timescales, which depending on their source, can range from hours, to milliseconds \cite{macintosh2005speckle}. The quasi-static nature of some of the speckles along with their resemblance with companion signals, poses a challenge for post-processing methods as it creates a noise floor that prevents observations from reaching the fundamental photon noise limit. Image processing has therefore been an active field of research since the mid-2000s, but no algorithm has brought detection limits down to the fundamental noise limit yet, except in the very specific case of space-based integral field spectroscopy\cite{Ruffio24}. 

To work towards this goal in ground-based observations, HCI relies on four main pillars: (i) adaptive optics, (ii) coronagraphy, and the relevant combination of (iii) an observing strategy and (iv) a post processing algorithm. Over the last two decades, the parallel development of these techniques has led to significant improvements in terms of achieved contrast through successive generations of AO-fed, coronagraphic HCI instruments. One such instrument is the Enhanced Resolution Imager and Spectrograph (ERIS) at the VLT, a second generation mid-infrared instrument with improved diffraction-limited HCI capabilities compared to its predecessor NACO \cite{davies2023enhanced}. It offers among others a vortex coronagraph, which is one of the most efficient coronagraph devices in terms of inner working angle (down to angular separations  of $\sim 1\,\lambda/D$), and its achromatic properties\cite{mawet2005annular,ORBi-9041e13f-528a-4e2d-ba46-e9b33d280a28}. Moreover, ERIS vortex coronagraphic mode shows an improvement of about 1 mag in detection limit compared to previous instruments such as NACO at the VLT\cite{de2024vlt}. In addition, the highly performing ERIS AO system provides access to its WFS telemetry (currently only in ``engineering mode'', but potentially generalised to general scientific operations in the future). 

\begin{figure}[t]
\begin{center}
\includegraphics[width=0.65\linewidth]{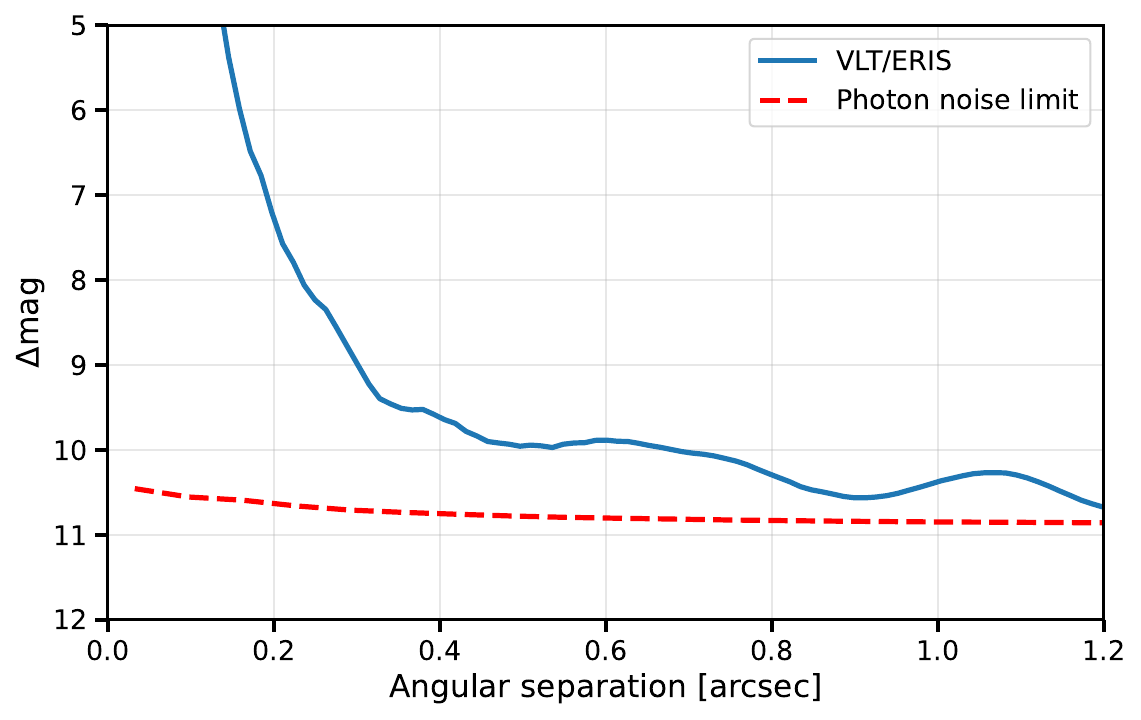}
\end{center}
\caption{\label{fig:noise_limit} $5\sigma$ post-processed contrast curve  of on-sky data obtain with the vortex focal plane coronagraph and median PSF subtraction (blue) acquired with ERIS (total ADI sequence of 70-min, \ang{59} rotation, magnitude in L band of $6.1$).
The theoretical noise limit is shown in red. As observed current post-processing methods remain far from this limit at close angular separations.}
\end{figure}

While these instrumental advances have helped to overcome the small separation and large flux ratio between a central star and the exoplanet signal of interest, speckle noise still limits observations. For this reason, differential observing techniques and post-processing algorithms have long been vital tools. For instance, Angular differential imaging (ADI) \cite{marois2006angular}, the most common observing technique in HCI, relies on the rotation of the field of view during the observation to differentiate between speckles (quasi-static) and astrophysical sources (rotating). Another common observing technique is reference-star differential imaging (RDI) \cite{mawet2012direct}, which uses the observation of a reference star to calibrate the coronagraphic point spread function (PSF) from the image. Equipped with these reference images, post-processing algorithms such as principal components analysis (PCA)\cite{amara2012pynpoint} are then able to subtract these modelled PSFs from the science images to reveal the companion signals. However, these current post-processing methods are sensitive to two major limitations: false positives due to the misclassification of speckles as exoplanet signals, and over-subtraction, where real signals from companions are mistakenly removed. As a result, current post-processing algorithms remain far from the photon noise limit at small angular separations. An example of this performance gap is shown in Fig.~\ref{fig:noise_limit}. 

\begin{figure}[t]
\begin{center}
\includegraphics[width=0.86\linewidth]{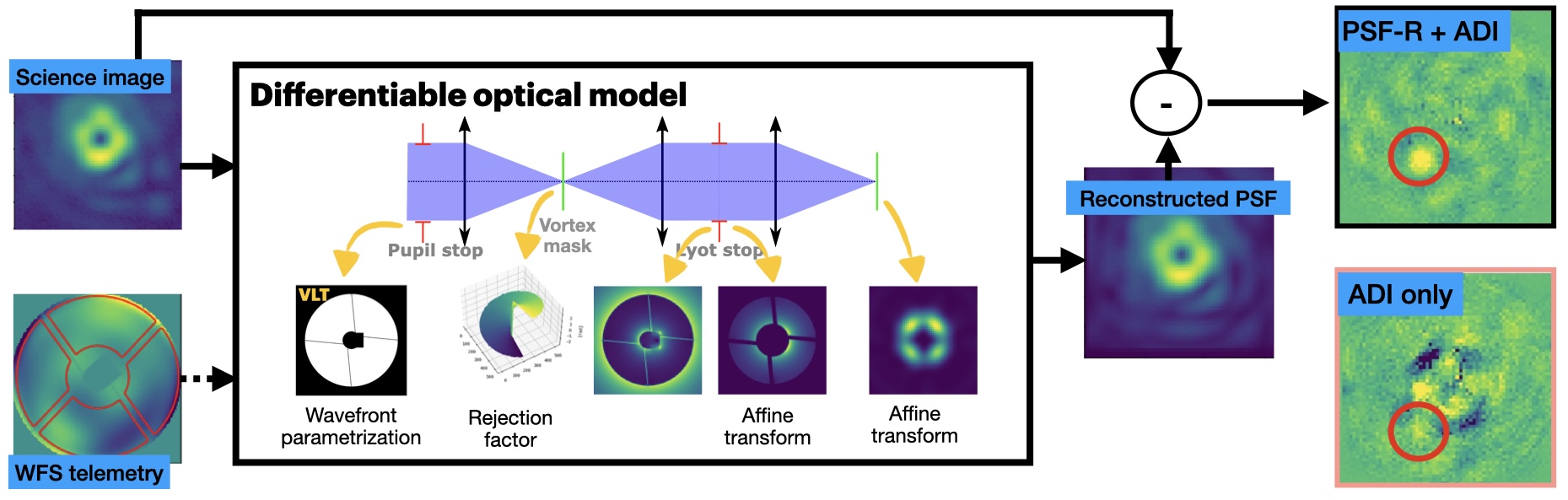}
\end{center}
\caption{\label{fig:diff_model} Illustration of our telemetry-and model-based post-processing principle for the VLT/ERIS instrument. The physical parameters of the model can be optimised via gradient-based optimisation to improve the reconstructed PSF. This reconstruction is then subtracted from the science observations to produce a clean, ideally speckle-free, image. The approach's potential is showcased against classical ADI by injecting an artificial companion (with $\Delta$mag=8 and located at \SI{0.2}{\arcsecond} from its star) into an ERIS dataset and comparing its recovery (see red circles). Our method strongly improves the significance of the detection. For practical scientific exploitation, WFS telemetry is essential to avoid model over-parameterisation and track the rapid wavefront temporal variations. }
\end{figure}

There is still therefore a large untapped potential in improving the sensitivity limit of high-contrast imaging observations. This motivates the development of methods that exploit information beyond the science images alone. In this context, we investigate the use of real-time WFS telemetry to enable PSF reconstruction. This approach, referred to as Wavefront-Assisted Reconstruction of the PSF (WARP), aims to improve starlight subtraction and increase sensitivity to faint exoplanets by explicitly modelling the wavefront aberrations that give rise to speckles, with a theoretical improvement of up to four magnitudes at the smallest angular separations according to Fig.~\ref{fig:noise_limit}. For its implementation we have developed a differentiable optical of the ERIS vortex coronagraph, illustrated in Fig.~\ref{fig:diff_model}. The model is parametrised by physically motivated variables, including entrance phase and amplitude aberrations, vortex retardance imperfection, and Lyot plane and detector plane misalignments. The differentiability allows gradient descent optimisation and precise model calibration. This produces a reconstructed PSF that we can then subtract from the on-sky images to reveal the incoherent signal. This method is inspired from a few recent works, most notably Guyon et al.\cite{guyon2022high} for the usefulness of AO telemetry in PSF reconstruction, Feng et al.\cite{feng2025exoplanet} for a first application of a differentiable optical model in a space-based HCI context, and Desdoigts et al.\cite{desdoigts2026amigo} for implementing an end-to-end, differentiable optical model calibrated via automatic differentiation with real on-sky data. 

As a first proof-of-concept, we injected an artificial planet in an ERIS dataset to illustrate the potential of a model-based PSF reconstruction and subtraction compared to a classical ADI post-processing technique (see Fig.~\ref{fig:diff_model}, right). While being particularly promising, this demonstration can hardly be used in practice due to the lack of ancillary data, hence necessitating an over-parametrisation of the entering wavefront. Including WFS telemetry, to capture the fast wavefront temporal variations, will alleviate this limitation and enable our approach to be scientifically exploited.

In this proceeding, we present our first results on telemetry-driven PSF reconstruction for the ERIS instrument. As an initial proof of concept, we investigate how accurately the on-sky PSF can be reproduced from wavefront sensor telemetry using a forward optical model.

\section{CHARACTERIZATION OF THE ERIS PSF}
\label{sec:jitter}
Differentiable optical models provide a powerful framework for reconstructing on-sky PSFs. However, before applying such a method, we first need to assess whether the WFS telemetry can accurately reproduce the PSF observed by the NIX science camera using a forward optical model. To this end, our analysis is based on a non-coronagraphic ERIS observation sequence. The sequence comprises science images, taken with a 40 ms integration time, along with their corresponding AO telemetry. The recorded telemetry has a loop sampling frequency of 1 kHz and consists of (i) wavefront sensor slopes measured by a Shack-Hartmann (SH) sensor, (ii) the control signals applied to the 1170 deformable mirror actuators, and (iii) AO calibration matrices. The telemetry was acquired in 60-sec batches at discrete intervals during the observation sequence. This study is based on one such telemetry snapshot, for which both the AO data and its corresponding science images are available. 

For this initial characterization of the ERIS PSF stability, we focused on the tip and tilt modes, which are directly measurable on the science images, making them the most direct metric to quantify image jitter and compare the AO telemetry with the science images. We did not consider higher-order aberrations in this initial analysis because of the difficulty to isolate their individual contributions. Although other low order aberrations, such as focus, are also affected by mechanical effects, we did not include them in this initial assessment.

Regarding the science images, the use of a non-coronagraphic observing sequence allowed us to accurately track the temporal evolution of the stellar PSF centroid. Before performing the described PSF centroid tracking, we applied the standard detector calibrations, including dark subtraction, flat field correction, and bad pixel masking. We obtained a first estimate of the stellar centroid in each frame using the centre of gravity (CoG) method. We then refined the centroid position by cross-correlating each frame with a reference image, obtained by median combining the frames at the same dither position. We performed this cross-correlation using the DFT upsampling method proposed in Ref.~\citenum{guizar2008efficient} and implemented in the VIP package \cite{gonzalez2017vip,christiaens2023vip}. This method estimates the cross-correlation peak using an FFT of the images  and then refines its location by upsampling the discrete Fourier transform around the estimated peak. We defined the reference centroid as the centroid of the median combined reference image. With this refinement, we were able to measure the PSF displacement relative to the reference centroid in the X and Y directions with sub-pixel precision. We then used the resulting time series to quantify the image jitter as recorded by the NIX science camera. 

To compare the jitter measurements with the AO telemetry, we reconstructed the first 20  modal coefficients from the recorded WFS data. We first assembled the Shack-Hartmann slope measurements from all subapertures into the slope vector, \textit{s}, and multiplied it by the calibrated slope-to-modes (\textit{S2M}) reconstruction matrix (provided by the ERIS AO calibration) to obtain the modal coefficients. We then retained only the tip and tilt coefficients for the present analysis and 
used them to reconstruct the corresponding phase aberrations. Finally, we propagated the reconstructed phase through a Fraunhofer model of the ERIS optical system, obtaining one mock PSF for each propagated wavefront. 

\begin{figure}[t]
\begin{center}
\includegraphics[width=0.7\linewidth]{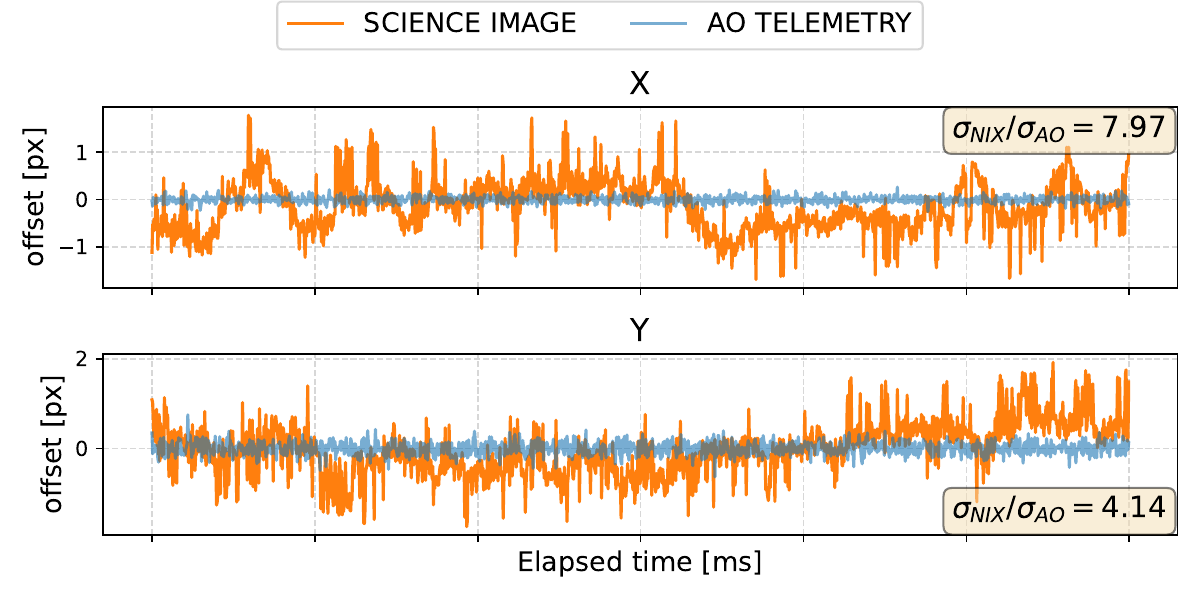}
\end{center}
\caption{\label{fig:jitter_comparison}
Comparison of the centroid offset time series measured from the NIX science images (orange) and reconstructed from the AO telemetry (blue). The centroid offsets are expressed in pixels (pixel scale: 13~mas), while the time axis is given in milliseconds. The top panel shows the X-axis centroid offsets, while the bottom panel shows the Y-axis centroid offsets.}
\end{figure}

As the AO telemetry was recorded at 1 kHz over a 60 s interval, we applied a decimation factor of 40 before propagating the reconstructed wavefront through the forward model of the instrument to match the frame rate of the ERIS-NIX science camera. Because of this, only every 40th wavefront was propagated, which yielded a PSF generation rate of 25 Hz. With this decimation, we reduced computational cost while preserving synchronisation with the science images. We then applied the same centroid pipeline described above to quantify the PSF jitter as predicted by the AO telemetry. Fig.~\ref{fig:jitter_comparison} shows the time series of the centroid offsets along the X and Y axes for the NIX PSF and the forward-modeled PSF from the AO telemetry. As observed, the time series reveal a discrepancy between the star jitter measured from the science images and that recorded by the AO telemetry.  The NIX jitter (orange) presents larger variability than the AO measurements (blue) in both axes, with RMS values of  $\sigma_{\mathrm{NIX}} = 7.1$ mas and $\sigma_{\mathrm{AO}} = 0.9$ mas in the X direction and  $\sigma_{\mathrm{NIX}} = 7.6$ mas and $\sigma_{\mathrm{AO}} = 1.8$ in the Y direction, corresponding to $\sigma_{\mathrm{NIX}}/\sigma_{\mathrm{AO}}$ ratios of $7.97$ and $4.14$, respectively. This mismatch suggests that the tip/tilt recorded by the AO telemetry is insufficient to fully explain the image motion observed in the science images.

\section{Evidence for Discrepancies in the Current Model}

In this section, we study whether the discrepancies identified in Section \ref{sec:jitter} propagate into the reconstructed PSF and hamper its ability to reproduce on-sky observations. To investigate this effect, we use an on-sky ERIS-NIX vortex coronagraph dataset. The observation used here was acquired in the Lp filter with an integration time of 10 ms per frame and a total duration of 60 s. Simultaneously, WFS telemetry was recorded throughout the sequence with a loop frequency of 1 kHz. This dataset enables a direct comparison between telemetry based PSF reconstruction and the measured NIX on-sky coronagraphic PSF,  as it provides synchronized images from the NIX science camera and AO telemetry. 

Only a subset of the science frames could be used since the stellar image was not consistently centred on the vortex phase mask. The vortex coronagraph phase mask performance has a high sensitivity to pointing drifts \cite{GOXivrySPIE}, which resulted in a significant fraction of the frames exhibiting large residual stellar leakage. To mitigate this effect, we used a lucky imaging approach that calculated the flux within a radius aperture of $1\,\lambda/D$  on each of the usable frames, and then we selected the frame that recorded the lowest flux, which corresponds to the best rejection factor of the vortex coronagraph. As a result, we were able to minimise the effect of residual stellar leakage caused by pointing drifts. The comparison we present here therefore evaluates whether the reconstructed PSF can reproduce the best-quality on-sky observation, based on this short dataset, rather than the temporal average of the sequence.

\begin{figure}[t]
\begin{center}
\includegraphics[width=0.75\linewidth]{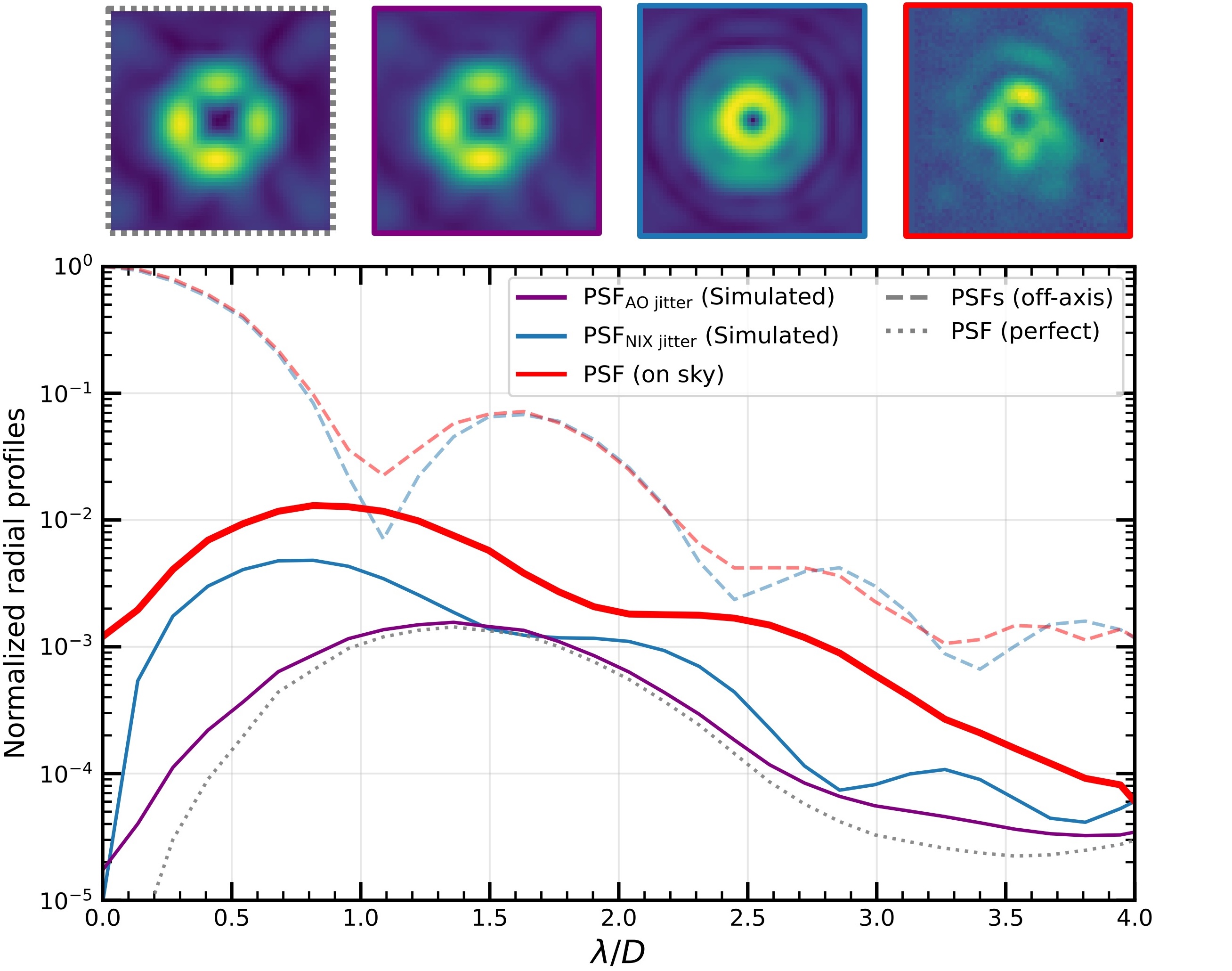}
\end{center}
\caption{\label{fig:rad_profile} (\textit{Top}) Modelled ERIS vortex PSFs and the on-sky PSF, ordered as follows: (1) PSF assuming a perfect, aberration-free wavefront, (2) PSF reconstructed from AO telemetry using Zernike modes up to the 20th order, (3) PSF reconstructed using higher-order aberrations up to the 20th Zernike mode with jitter measured by the NIX science camera and (4) the on-sky PSF. (\textit{Bottom}) Normalised radial intensity profiles with the star centred on the vortex phase mask (solid lines) and off-axis (dashed lines) as a function of angular separation. The on-sky PSF is shown in red, the PSF modelled from telemetry reconstruction (up to 20th Zernike mode) in purple , the PSF reconstructed from the NIX camera tip-tilt measurements and AO higher-order modes in blue, and the simulated PSF assuming a perfect wavefront in grey. As illustrated, the telemetry based profile does not match the on-sky vortex observations and bears a stronger resemblance to the unaberrated profile. In contrast, the NIX tip-tilt profile aligns more closely with the on-sky observations. 
}
\end{figure}

We now focus on the AO-derived PSFs. To reconstruct the coronagraphic PSFs from the telemetry data, we followed the modal reconstruction procedure described in Section~\ref{sec:jitter}. In addition to the tip and tilt modes, we included higher-order aberration coefficients recovered from the AO telemetry up to the 20th Zernike mode. We propagated the reconstructed modes through a Fraunhofer model of the ERIS optical system, including a charge-2 vector vortex coronagraph. Using this framework, we generated three modelled coronagraphic PSFs: (i) simulated PSF assuming a perfect wavefront, (ii) simulated PSF reconstructed solely with AO telemetry and (iii) simulated PSF that combines the tip-tilt measurements from the NIX science camera described in Section~\ref{sec:jitter} with higher order aberrations reconstructed from the AO telemetry. We then compared these three PSFs with the selected on-sky science frame to assess the effect of the jitter discrepancy reported in Section~\ref{sec:jitter} on the final result. Figure~\ref{fig:rad_profile} (top) displays these three modelled PSFs and the on-sky frame.  To study their agreement, we calculated the azimuthally averaged radial intensity profiles of each PSF using the \texttt{frame\_average\_radprofile} routine from the VIP package\cite{gonzalez2017vip,christiaens2023vip}. All profiles were normalized by the peak intensity of their corresponding off-axis PSFs.

Figure \ref{fig:rad_profile} (bottom) illustrates these profiles as a function of angular separation ($\lambda/D$). The PSF reconstructed using only AO telemetry (purple) shows a strong discrepancy with respect to the on-sky vortex PSF (red), reaching almost two orders of magnitude at the shortest angular separations. The telemetry-derived profile has a much better agreement with the perfect PSF profile (dotted grey curve) than with the on-sky observations. This suggests that the underestimate of the residual jitter we previously identified in Section \ref{sec:jitter} propagates into the reconstructed PSF, limiting its ability to reproduce the on-sky coronagraphic intensity profile. The reconstructed PSF derived from  NIX tip-tilt measurements (blue) shows better (though not perfect) agreement with the on-sky vortex coronagraph data. This result provides further evidence that mechanical effects affecting the tip and tilt modes are being unaccounted for in the AO telemetry. 

Despite this improvement, some discrepancies remain between the reconstructed PSF (blue curve) and the on-sky measurements (red curve). One possible explanation is that mechanical disturbances are not limited to tip and tilt but that they are also affecting other low order modes such as defocus, which in this study we have reconstructed solely from the AO telemetry. Alternatively, the remaining discrepancies may arise due to the fact that residual image jitter estimated from the non-coronagraphic observations is not representative of the coronagraphic dataset. This jitter would need to be at least 4 times larger than that measured for the non-coronagraphic dataset for the simulated profile to reach a level compatible with the on-sky data, and even then it would not fully explain the observed profile. This indicates that additional effects, such as Lyot stop misalignments, residual NCPAs, or other unmodelled instrumental effects, are also contributing to the discrepancy. 

\section{Conclusion and Future Work}

In this contribution, we identified a discrepancy between the ERIS vortex coronagraph PSF reconstructed from adaptive optics telemetry and the PSF observed in on-sky data. Our analysis shows that the jitter measured by the NIX science camera has a larger standard deviations than those recorded by the AO telemetry, by factors up to 8, depending on the direction. Reconstructing the PSF using the science camera jitter derived from another (non-coronagraphic) data set achieves better agreement with the observed on-sky PSF than using the AO telemetry alone. These results suggest that the discrepancy is likely caused by mechanical effects that contribute to image jitter in the NIX science camera. Although this interpretation remains to be confirmed, it provides a reasonable explanation for the behaviour we observed on the reconstructed PSFs. 

Motivated by these findings, future work will focus in parallel on mitigating the impact of the observed jitter and improving its characterization. In particular, we will investigate whether the residual jitter can be parameterised and incorporated as an additional parameter in our differentiable optical model of ERIS, while also identifying the instrumental source responsible for the recorder jitter. 

To get real time measurements of the stellar tip and tilt behind the vortex, we propose implementing satellite spots, following the approach adopted by other instruments like SPHERE \cite{beuzit2019sphere}. These consist of four diffraction spots generated during the observing sequence by applying a waffle pattern to the deformable mirror. The satellite spots provide a reliable reference to monitor jitter even for an occulted star and would allow us to accurately estimate the tip and tilt error in real time. 

As a separate study, we also propose investigating the origin of the observed jitter. Since the cryogenic system of ERIS is suspected to be a major source of mechanical vibrations, we propose to repeat the data acquisition process with the cryogenic pump switched off. Comparing the two datasets will allow us to quantify the impact of the cryogenic pump on the image jitter measured on the science camera and explore possible mitigation strategies. Finding a way to mitigate the source of unseen tip-tilt jitter would not only reduce speckle noise, but would also bring down the fundamental photon noise limit by decreasing the average stellar residual level in the recorded data, and should therefore be considered a priority for ERIS.

Looking ahead, the use of a differentiable optical model for ERIS PSF reconstruction will depend critically on the accuracy of the input AO telemetry. In this proceeding we showed that the current telemetry underestimates the jitter recorded by the science camera, limiting the reproducibility of the on-sky vortex PSF from WFS information alone. For future works we will therefore focus on accounting for these discrepancies, either by identifying their physical origin or by parameterising their effects so that they can be included in our differentiable optical model of the ERIS instrument, which is the final goal of this study. 

\acknowledgments 

On top of \texttt{VIP}\cite{gonzalez2017vip,christiaens2023vip} and \texttt{eristools} (Bonse et al. 2026, in prep.), this work also makes use of \texttt{aotpy}\cite{Gomes+24}.
GOX thanks Alfio Puglisi and Armando Riccardi for useful discussions on the ERIS AO telemetry.

\bibliography{report} 
\bibliographystyle{spiebib} 

\end{document}